\documentclass[review]{elsarticle}
\journal{Computer Physics Communications}

\def\ack{\section*{Acknowledgements}%
  \addtocontents{toc}{\protect\vspace{6pt}}%
  \addcontentsline{toc}{section}{Acknowledgements}%
}
\usepackage{lmodern}
\usepackage{lmodern}
\usepackage[T1]{fontenc}
\usepackage{amsmath}
\usepackage{amssymb}
\usepackage{graphicx}
\usepackage{esint}
\biboptions{numbers,sort&compress}
\makeatletter

\usepackage{amsthm}\usepackage{latexsym}\usepackage{bm}\usepackage{amsfonts}
\usepackage{hyperref}

\usepackage{listings}
\usepackage{xcolor}
\begin{document}
\begin{frontmatter}
\title{The Accurate Particle Tracer Code}

\author[address1]{Yulei Wang}
\author[address1]{Jian Liu\corref{corresponding}}
\cortext[corresponding]{Corresponding author}
\ead{jliuphy@ustc.edu.cn}
\author[address1,address2]{Hong Qin}
\author[address3]{Zhi Yu}
\author[address1]{Yicun Yao}

\address[address1]{School of Nuclear Science and Technology and Department of Modern
Physics, University of Science and Technology of China, Hefei, Anhui
230026, China}
\address[address2]{Plasma Physics Laboratory, Princeton University, Princeton, NJ 08543, USA}
\address[address3]{Theory and Simulation Division, Institute of Plasma Physics, Chinese
Academy of Sciences, Hefei, Anhui 230031, China}

\begin{abstract}
The Accurate Particle Tracer (APT) code is designed for systematic
large-scale applications of geometric algorithms for particle dynamical
simulations. Based on a large variety of advanced geometric algorithms,
APT possesses long-term numerical accuracy and stability, which are
critical for solving multi-scale and nonlinear problems. To provide
a flexible and convenient I/O interface, the libraries of Lua and
Hdf5 are used. Following a three-step procedure, users can efficiently
extend the libraries of electromagnetic configurations, external non-electromagnetic
forces, particle pushers, and initialization approaches by use of
the extendible module. APT has been used in simulations of key physical
problems, such as runaway electrons in tokamaks and energetic particles
in Van Allen belt. As an important realization, the APT-SW version
has been successfully distributed on the world's fastest computer,
the Sunway TaihuLight supercomputer, by supporting master-slave architecture
of Sunway many-core processors. Based on large-scale simulations of
a runaway beam under parameters of the ITER tokamak, it is revealed
that the magnetic ripple field can disperse the pitch-angle distribution
significantly and improve the confinement of energetic runaway beam
on the same time. 
\end{abstract}
%%\maketitle
\begin{keyword}
\texttt{}Structure-Preserving Algorithms\sep Plasma simulation \sep Multi-timescale Dynamics \sep Large-scale Simulation 
\end{keyword}

\end{frontmatter}
%%\linenumbers

\section{Introduction\label{sec:Introduction}}

Nonlinear and multi-scale dynamical processes are ubiquitous in different
fields of scientific and engineering researches. Especially in plasma
physics, where long-range collective phenomena dominate, advanced
numerical schemes and powerful computing software are required for
solving complex physical and technical problems. The GeoAlgorithmic
Plasma Simulator (GAPS) project is initiated in order to solve various
difficult yet key problems in plasma-related domains by applying advanced
geometric algorithms and modern large-scale simulation techniques.
The Accurate Particle Tracer (APT) code is one product of the GAPS
project and aims to facilitate the systematic large-scale applications
of advanced algorithms for particle dynamical simulations.

Particle dynamical simulations play important roles in numerical studies
of plasmas. Various codes have been developed and used widely for
solving problems including the particle acceleration in magnetic reconnection
\cite{Perona_Eriksson_testPtcCode2014,Hamilton_TestParticleCode_CUEBIT_MHD_rcnct},
fast particle dynamics in Tokamaks \cite{Pfefferle_VENUS-LEVIS_2014,White_ORBIT_1984},
wave-particle interactions in earth magnetosphere \cite{SuZhenPeng_2014_EMIC},
and particle transportation in plasma turbulences \cite{Dalena_2012_testparticle,Tautz_2010}.
The particle dynamical simulation is also a main method of solving
differential stochastic equations, when stochastic processes of plasmas
are considered, such as the simulation of runaway dynamics during
disruptions of tokamak operations \cite{Eriksson_Helander_SimRE_disruption_2003}. 

Although fruitful results have been achieved, the accumulation of
numerical errors makes traditional codes unreliable after long-term
simulations which are unavoidable for addressing multi-scale and nonlinear
problems. Real physical information may also be distorted due to the
breakdown of original physical structures in computation schemes.
To guarantee long-term numerical accuracy and stability, a series
of advanced geometric algorithms have been systematically developed
recently \cite{McLachlan_GeoAlgrithm_background,Candy_SympAlg_SepHam_1991,Qin_VariatianalSymlectic_2008,CSPIC_2016,Qin_Boris_2013,Guan_Qin_Sympletic_RE,Ruili_VPA_2015,Jianyuan_Multi_sympectic_2013,HeYang_HamiltonTimeInt_VMs_2015,Ruili_GC_canonical_2014,HeYang_HigherOrderVPA_2016,HeYang_Spliting_2015,Channell_1990_sympInteger,Channell_Accelerator2014,HeYang_Ksymp_PLA_2016,Josh_POP_2012,Kraus_Varianal4Nonvariational2015,Kraus_VariationalSym_Thesis,Lee_Qin_PPCF_2015,LiJinXing_GC_Symp_2011,McLachlan_AccuracyOfsymInt1992,Shadwick_Variational_2014,ZhangRuili_ExpGenerateSym_2016,ZhangRuili_ExpGenerateSym_Rel_2016,ZhuBeiBei_Ksym_2016,XiaoJY_PIC_wave_2015},
which bound the global numerical errors by preserving geometric structures.
On the other hand, the further development and application of geometric
algorithms needs interdisciplinary cooperation. An efficient platform
for researchers in different fields is needed to integrate the latest
trans-disciplinary achievements. 

In this paper, we introduce the design details, numerical strategies,
use instructions, and novel physical results of the APT code. Different
from the particle dynamical codes equipped with traditional algorithms
and focusing only on some specific physical application scenarios
\cite{Perona_Eriksson_testPtcCode2014,Hamilton_TestParticleCode_CUEBIT_MHD_rcnct,Pfefferle_VENUS-LEVIS_2014,Dalena_2012_testparticle,Tautz_2010},
the APT code is based on geometric algorithms with secular stability,
and its algorithm and physical libraries can also be extended conveniently
to accomplish tasks in various research fields. Therefore, APT not
only can be used to address multi-scale and nonlinear problems, but
also serves as a universal platform for researchers from different
fields, such as plasma physics, accelerator physics, space science,
fusion energy research, computational mathematics, software engineering,
and high-performance computation.

The underlying model of APT is the first principle particle dynamical
equations. In simulations of plasma systems, the distribution of plasmas
is sampled statistically in the phase space. Using appropriate geometric
algorithms, APT traces each sample point accurately by solving full-orbit
Lorentz force equations that are the characteristic line equations
of the Vlasov equation. If considering the random collisional terms,
the Langevin approach is used, which transfers the collisional terms
to random forces \cite{Cadjan_LangevinCoulombCollision_1999}. The
external electromagnetic fields are set up through analytical functions,
or discrete field configuration data obtained from experiments or
simulations. There already exist many built-in electromagnetic field
configurations for typical physical problems in APT. Self-consistent
calculation of APT is implemented through dynamical field models,
namely, field functions with variable parameters evolving together
with the phase-space states of all particles. 

The APT code is implemented in standard C language and can be distributed
directly on Unix-like operation system. For non-Unix-like systems,
users can build APT via Unix-like compatibility layers, such as Cygwin
and MSYS. The APT code consists of seven main modules, including the
I/O module, the initialization module, the particle pusher module,
the parallelization module, the field configuration module, the external
force-field module, and the extendible module. The I/O module calls
the libraries from Lua and Hdf5 projects. The input configuration
files consist of several Lua scripts which make it convenient and
flexible for users to set the parameters of physical problems. The
output data is stored in the Hdf5 format, which enables users to access
data in a file-system-like way and simplifies data analysis. The initialization
module provides a number of methods for statistical sampling. The
particle pusher module contains various advanced geometric algorithms,
including volume-preserving and symplectic integrators with different
orders and stability domains. Traditional algorithms such as Runge-Kutta
method with different orders are also available. It is thus convenient
to choose appropriate algorithms for solving realistic problems and
studying numerical methods under complex physical setup. To boost
large-scale applications of geometric algorithms, the parallelization
module of the stable APT distribution supports the MPI parallelization.
Another version of APT supporting CUDA is also developed. As an important
parallel version, the branch version APT-SW has been built recently
for the world's fastest supercomputer, the Sunway TaihuLight supercomputer
\cite{Top500_web}, which provides more than ten million computation
cores and has the peak performance of 125PFlops \cite{TaihuLight_2016}.
To fully utilize the computation ability of the Sunway TaihuLight,
APT-SW is designed to support the Sunway many-core parallelization.
Supported by the powerful computation capacity, APT-SW can be used
to simulate processes with large amounts of sampling particles, complex
field configurations, and multi-timescales spanning more than 10 orders.
The field configuration module and the external force-field module
contain respectively various electromagnetic configurations and other
external non-electromagnetic force fields, such as radiation force,
collisional force, bremsstrahlung force, and gravitation field. Written
by the bash-script, the extendible module provides a convenient way
to extend the source code. Following the interfaces of APT and a three-step
extending procedure, one can easily add new configuration parameters,
field configurations, external forces, algorithms, and initialization
functions into APT. These new packages form a new version of APT,
and can also be used by other researchers. The detailed introductions
about the interfaces and the procedure to use and extend APT are provided
in this paper.

APT has been used in the study of geometric numerical methods and
the simulations of some multi-scale physical processes \cite{CollisionlessScater_NF_Letter_2016,RELong_WangYulei2016}.
In this paper, we present two benchmarks from plasma physics. In the
first case, the multi-scale dynamics of runaway electrons in tokamaks
is studied by secular simulations over $10^{12}$ time steps. Using
the relativistic volume-preserving algorithm in APT, the refined structures
of runaway transit orbits are revealed. The APT code shows long-term
numerical accuracy and enables the discovery of novel mechanisms such
as neoclassical collisionless pitch-angle scattering and new runaway
energy limit rule \cite{CollisionlessScater_NF_Letter_2016,RELong_WangYulei2016}.
In the second case, the evolution of energetic particle distribution
in Van Allen belt is simulated by tracing massive sampling particles.
The symplectic Euler pusher for relativistic dynamics is employed.
The evolution of energetic particles in the terrestrial magnetic field
is recovered precisely. 

New physical results have been obtained based on the large-scale simulations
of APT-SW. The long-term evolution of the runaway electron beam under
the realistic parameters of ITER tokamak is simulated on the TaihuLight
supercomputer. This simulation involves approximately $10^{7}$ samplings,
$10^{11}$ iteration steps, and more than $10^{21}$ floating-point
operations, which is currently the largest particle simulation. Both
the neoclassical collisionless pitch-angle scattering \cite{CollisionlessScater_NF_Letter_2016}
and the magnetic ripple stochastic effect are considered \cite{RELong_WangYulei2016,Rax_RE_resonance1993}.
Without magnetic ripple field, after long-term evolution, the pitch-angle
distribution tends to be concentrated in a small interval and has
non-zero average value, which is consistent with the results in Ref.\,\cite{RELong_WangYulei2016}.
However, if the magnetic ripple is considered, the pitch-angle distribution
can be significantly dispersed. Meanwhile, the poloidal profile of
runaway beam is proven to be squeezed towards tokamak core by the
magnetic ripple field, which serves as an evidence to the improved
confinement of runaway electrons. 

The remaining part of the paper is organized as follows. The introductions
on APT, including the numerical strategy, the parallelization methods,
and the procedures of using and extending APT, are provided in Sec.\,\ref{sec:Design}.
In Sec.\,\ref{sec:Benchmarks}, two typical benchmarks under different
physical backgrounds carried out by APT are exhibited. The novel physical
results of large-scale simulation of runaway electrons on the Sunway
TaihuLight supercomputer are discussed in Sec.\,\ref{sec:Taihu_results}.
Section \ref{sec:Conclustion} gives a brief summary.

\section{Architecture of Apt\label{sec:Design}}

\begin{figure}
\includegraphics[scale=0.25]{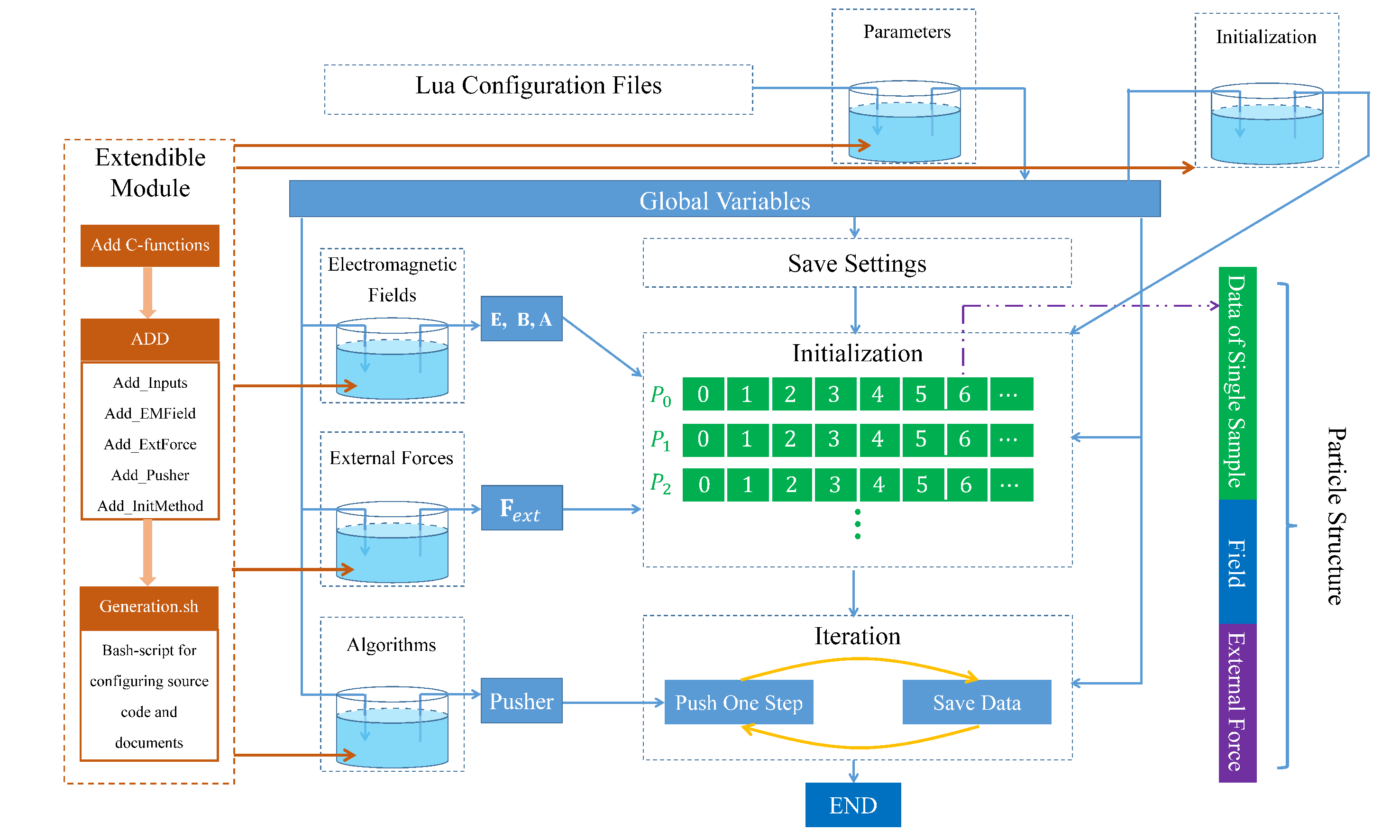}\caption{Diagram of the APT architecture. The blue arrows denote the flow of
procedure, while the brown parts represent the extendible module.
In the Initialization box, sampling particles assigned to different
processors are marked by $P_{0}$, $P_{1}$, $P_{2}$, $\cdots$.\label{fig:APT_Arch}}
\end{figure}

As a product of the GAPS project, APT aims to benefit simulations
of various research fields through the long-term accuracy of geometric
algorithms. In order to accelerate the development and optimization
of geometric algorithms, and form a convenient platform for integrating
novel techniques from different fields, the architecture of APT is
designed to be modularized and extendible. Figure \ref{fig:APT_Arch}
depicts the design schematic of the APT code. There are five containers
for configuration parameters, electromagnetic functions, external
forces, particle pushers, and initialization methods. Besides using
the functions already integrated in APT, one can also conveniently
add new variables, electromagnetic field configurations, external
forces, algorithms as well as initialization approaches into these
containers through the extendible module, see the brown arrows in
Fig.\,\ref{fig:APT_Arch}. The built-in geometric algorithms in the
pusher container provide the APT code with secular stability and accuracy.
As is shown by the blue arrows in Fig.\,\ref{fig:APT_Arch}, users
can set all configuration parameters required by a specific task in
one Lua configuration file, which loads appropriate values into the
parameter container. Then, these parameters control APT to load selected
modules from corresponding containers, and manage the processes of
parallelization, iteration, and data output. In this section, we introduce
the numerical strategy and the parallelization implementations of
APT. Meanwhile, we exhibit the detailed procedures of installing,
using, and extending APT. 

\subsection{The numerical strategy of APT\label{sub:GeoAlgKer}}

The fundamental numerical method of the APT code is based on the first
principle particle dynamical equations. The evolution of every particle
is governed by both the electromagnetic force (the Lorentz force)
and external non-electromagnetic forces. When dealing with the electromagnetic
forces, APT uses structure-preserving geometric algorithms to ensure
the secular numerical stability. Because the design of APT aims to
form a platform for studying and using geometric algorithms, there
is more than one geometric pusher integrated in the particle pusher
module. If more users join the APT project and extend the libraries,
more algorithms will be included. So, it is not possible and not necessary
to present the details of all built-in algorithms. However, it is
still helpful to exhibit two typical algorithms to clarify the framework
of numerical strategy used in APT.

The first example is the volume-preserving algorithm (VPA) for relativistic
charged particles, which solves directly the Newton motion equation,
namely,
\begin{eqnarray}
\frac{\mathrm{d}\mathbf{x}}{\mathrm{d}t} & = & \frac{\mathbf{p}}{\mathrm{m}\gamma}\,,\label{eq:Lorentz3_x}\\
\frac{\mathrm{d}\mathbf{p}}{\mathrm{d}t} & = & q\left(\mathbf{E}+\frac{\mathbf{p}\times\mathbf{B}}{\mathrm{m}\gamma}\right)\,,\label{eq:Lorentz3_p}
\end{eqnarray}
where $\mathbf{x}$ and $\mathbf{p}$ denote respectively position
and momentum, $\gamma$ is the Lorentz factor, $\mathrm{m}$ is rest
mass of a particle, $q$ is the particle charge, and $\mathbf{E}$
and $\mathbf{B}$ denote electric field and magnetic field, respectively.
Various kinds of VPAs have been established in different ways \cite{Qin_Boris_2013,Ruili_VPA_2015,HeYang_HigherOrderVPA_2016,HeYang_Spliting_2015}.
For example, the discrete equations of a 2-order VPA integrated in
APT is
\begin{eqnarray}
\mathbf{x}_{k+\frac{1}{2}} & = & \mathbf{x}_{k}+\frac{\Delta t}{2}\frac{\mathbf{p}_{k}}{\sqrt{\mathrm{m}^{2}+\mathbf{p}_{k}^{2}/c^{2}}}\,,\label{eq:VPA_1}\\
\mathbf{p}^{-} & = & \mathbf{p}_{k}+\frac{\Delta t}{2}\left(q\mathbf{E}_{k+\frac{1}{2}}\right)\,,\label{eq:VPA_2}\\
\mathbf{p}^{+} & = & \mathrm{Cay}\left(\frac{q\Delta t\hat{\mathbf{B}}_{k+1/2}}{2\sqrt{\mathrm{m}^{2}+\left(\mathbf{p^{-}}\right)^{2}/\mathrm{c}^{2}}}\right)\mathbf{p}^{-}\,,\label{eq:VPA_3}\\
\mathbf{p}_{k+1} & = & \mathbf{p}^{+}+\frac{\Delta t}{2}\left(q\mathbf{E}_{k+\frac{1}{2}}\right)\,,\label{eq:VPA_4}\\
\mathbf{x}_{k} & = & \mathbf{x}_{k+\frac{1}{2}}+\frac{\Delta t}{2}\frac{\mathbf{p}_{k+1}}{\sqrt{\mathrm{m}^{2}+\mathbf{p}_{k+1}^{2}/c^{2}}}\,,\label{eq:VPA_5}
\end{eqnarray}
where the subscript, $k$, denotes the $k$-th step, $\Delta t$ is
the time step length, $\hat{\mathbf{B}}$ is defined as
\begin{equation}
\hat{\mathbf{B}}=\left(\begin{array}{ccc}
0 & B_{z} & -B_{y}\\
-B_{z} & 0 & B_{x}\\
B_{y} & -B_{x} & 0
\end{array}\right)\,,\label{eq:Bhat}
\end{equation}
and the symbol $\mathrm{Cay}$ denotes the Cayley transform \cite{Ruili_VPA_2015}.
Another example is 1-order canonical symplectic Euler algorithm for
relativistic particles, which discretizes the Hamiltonian equation,
\begin{equation}
\frac{\mathrm{d}\mathbf{P}}{\mathrm{d}t}=-\frac{\partial H}{\partial\mathbf{X}}\,,\qquad\frac{\mathrm{d}\mathbf{X}}{\mathrm{d}t}=-\frac{\partial H}{\partial\mathbf{P}}\,,\label{eq:Hamiltonial}
\end{equation}
where, $\mathbf{P}$ and $\mathbf{X}$ are respectively canonical
momentum and canonical coordinate, $\mathbf{A}$ denotes the vector
potential, $\phi$ is the scalar potential, and the Hamiltonian is
$H\left(\mathbf{P},\mathbf{X}\right)=\sqrt{\left[\mathbf{P}-q\mathbf{A}\left(\mathbf{X}\right)\right]^{2}\mathrm{c}^{2}+\mathrm{m^{2}c^{4}}}+q\phi$.
The corresponding canonical symplectic Euler algorithm is
\begin{eqnarray}
\mathbf{P}_{k+1} & = & \mathbf{P}_{k}-\Delta t\left[-\frac{q\left(\mathbf{P}_{k+1}-q\mathbf{A}_{k}\right)\cdot\nabla\mathbf{A}_{k}}{\mathrm{m}\gamma\left(\mathbf{P}_{k+1},\mathbf{X}_{k}\right)}+q\nabla\phi_{k}\right]\,,\label{eq:symEulerP}\\
\mathbf{X}_{k+1} & = & \mathbf{X}_{k}+\Delta t\left[\frac{\mathbf{P}_{k+1}-q\mathbf{A}_{k}}{\mathrm{m}\gamma\left(\mathbf{P}_{k+1},\mathbf{X}_{k}\right)}\right]\,.\label{eq:symEulerX}
\end{eqnarray}

For charged particle systems with electromagnetic forces, the geometric
structures and corresponding structure-preserving algorithms have
been studied in detail. Fruitful results have been published on this
topic. However, in many situations, physical systems can be very complex,
and involve many non-electromagnetic forces. Some of these forces
possess clear mathematical structures, such as gravitational force,
while some don't, such as radiation force of charged particles. When
dealing with external non-electromagnetic forces, for geometric algorithms
based on Newton equations like Eqs.\,\ref{eq:VPA_1}-\ref{eq:VPA_5},
the basic idea of APT is to separate the calculations of electromagnetic
and non-electromagnetic forces. For the default calculation mode,
the electromagnetic forces of motion equation are calculated by structure-preserving
algorithms, while the non-electromagnetic forces, $\mathbf{F}_{ext}$,
are treated as effective electric fields within the discrete one-step
map. To be specific, after including the external force terms, Eq.\,\ref{eq:VPA_2}
and Eq.\,\ref{eq:VPA_4} become
\begin{eqnarray}
\mathbf{p}^{-} & = & \mathbf{p}_{k}+\frac{\Delta t}{2}\left(q\mathbf{E}_{k+\frac{1}{2}}+\mathbf{F}_{ext,k+\frac{1}{2}}\right)\,,\label{eq:VPA_2-1}\\
\mathbf{p}_{k+1} & = & \mathbf{p}^{+}+\frac{\Delta t}{2}\left(q\mathbf{E}_{k+\frac{1}{2}}+\mathbf{F}_{ext,k+\frac{1}{2}}\right)\,.\label{eq:VPA_4-1}
\end{eqnarray}
Generally speaking, the external forces can be functions of arbitrary
order derivatives of $\mathbf{x}$, namely, $\mathbf{F}_{ext}=\mathbf{F}_{ext}\left(\mathbf{x},\dot{\mathbf{x}},\ddot{\mathbf{x}},\cdots\right)$.
When treating these forces as effective electric fields, within one-step
map, the values of all the derivatives of $\mathbf{x}$ are calculated
at k-th step. Besides the default algorithms of calculating non-electromagnetic
forces, the extendible module of APT provides interfaces of the external
forces for users, which can help them design their own pushers for
addressing complex physical situations. Most of symplectic methods,
on the other hand, are obtained through discrete Hamiltonian or Lagrange
systems. Unlike the Newton equations, the non-electromagnetic forces
cannot be separated easily from Hamiltonian or Lagrange equations.
Under current version of APT, the external force module cannot be
used for such algorithms. 

Another important part of the numerical strategy is the calculation
of the electromagnetic field. In APT, the electromagnetic fields can
be calculated via analytic functions or discrete field data. The analytic
field functions include $\mathbf{E}\left(\mathbf{x},t\right)$, $\mathbf{B}\left(\mathbf{x},t\right)$,
$\mathbf{A}\left(\mathbf{x},t\right)$, $\phi\left(\mathbf{x},t\right)$,
and their derivatives, such as $\partial_{i}A^{j}\left(\mathbf{x},t\right)$,
$\partial_{i}\phi\left(\mathbf{x},t\right)$, $\partial_{i}\partial_{j}A^{k}\left(\mathbf{x},t\right)$,
$\partial_{i}\partial_{j}\phi\left(\mathbf{x},t\right)$, $\cdots$.
Self-consistent calculation of APT is implemented through dynamical
field models, which evolve under the control of parameters calculated
from the phase-space state of all particles. For example, the self-consistent
magnetic field of a charged beam is calculated based on line current
model. APT first gathers the phase-space information of all particles,
and then summarizes the information to obtain the evolution parameters
of self-consistent field, such as the position and the strength of
the current, and finally updates field models through new parameters.
The general model for self-consistent fields can be summarized as
$\mathbf{F}\left[t,\mathbf{x},c_{1}\left(\mathbf{x}_{i},\mathbf{p}_{i}\right),c_{2}\left(\mathbf{x}_{i},\mathbf{p}_{i}\right),\cdots\right]$,
where the index $i$ covers all the simulated particles, and $c_{1}$
and $c_{2}$ are dynamical parameters. Discrete field data from experiments
or simulations can also be used to calculate fields. APT accepts discrete
field data distributed on 3-dimentional cubic grids. The local fields
of particles are calculated via interpolation functions. To be specific,
for discrete vector potential field data $A_{ijk}^{l}$, the field
at $\mathbf{x}$ is given by $A^{l}\left(\mathbf{x}\right)=\sum_{ijk}A_{ijk}^{l}W\left(\mathbf{x}-\mathbf{x}_{J}\right)$.
Here $W$ is weighting function, $i$, $j$, $k$ are indexes on three
directions, and $l$ denotes the components of $\mathbf{A}_{ijk}$.

\subsection{The parallelization module\label{sub:ParaMod}}

\subsubsection{Overview\label{sub:Para_overview}}

In order to boost the large-scale application of geometric algorithms,
the parallelization module of APT is designed to support parallel
computation on different hardware architectures. On supercomputers
with Intel architecture, the APT code can be distributed via standard
MPI libraries. The GPU version of APT supports the CUDA parallelization
method. As a special implementation, APT-SW supports the many-core
acceleration of the Sunway TaihuLight supercomputer. The parallelization
module assigns the particle samples to different cores and works in
three optional modes, namely, the asynchronous mode, the synchronous
mode, and the quasi-synchronous mode. The selection of parallelization
modes is determined by specific physical models. The asynchronous
parallelization simulates particles independently and is applied to
problems with ignorable self-consistent fields of particles. This
mode minimizes the communications among processors, and thus possesses
highest efficiency among the three methods. For cases with the self-consistent
fields, the synchronous parallelization is required. In this mode,
particle samples are pushed synchronously and phase-space states of
all samples are collected for every step to update self-consistent
fields. The quasi-synchronous mode is developed to reduce the resource
wastes resulting from communications and is mainly designed for a
certain type of physical processes on large-scale computers, where
the timescale of self-consistent fields is larger than the timescale
of particle motion. For example, during the evolution of runaway beams
in tokamaks, the characteristic timescale of Lorentz force is on the
order of $10^{-11}\,\mathrm{s}$, while the timescale of self-consistent
field is on the order of $10^{-1}\,\mathrm{s}$. Instead of updating
the self-consistent field at every step, the quasi-synchronous model
only collects global information and updates the fields at some specified
moments, which are determined by the timescale of self-consistent
fields and the total physical duration of the simulation. Consequently,
the total number of communications is much less than calculation operations.
The communication bottleneck can thus be eased when one is dealing
with multi-scale self-consistent problems.

\subsubsection{APT-SW on the Sunway TaihuLight supercomputer\label{sub:Para_Taihu}}

Recently, the Top 1 supercomputer in the world has been updated as
the construction of the Sunway TaihuLight supercomputer in Wuxi, China.
Possessing peak performance of 125PFlops, the Sunway TaihuLight becomes
the fastest supercomputer in the world \cite{Top500_web,TaihuLight_2016}.
The calculation capability of the Sunway TaihuLight is provided by
$40000$ SW26010 many-core processors, each of which has 260 processing
elements \cite{TaihuLight_2016}. Therefore, there are more than $10^{7}$
cores available on this supercomputer. The Sunway TaihuLight supercomputer
system has a total storage of 20PB, which makes it possible for the
storage and analysis of big data. However, it is not straightforward
to distribute a program on this supercomputer, because the architecture
of the processor is different from any other CPUs or GPUs. In each
SW26010 processor, four master cores and 256 slave cores are integrated.
The two types of cores have similar computation ability but different
architectures. A program should utilize both the master and slave
cores to avoid the waste of the calculation capability of the Sunway
TaihuLight. Meanwhile, the communication between a master core and
its slave cores becomes the bottleneck of efficiency because of the
slow speed of accessing shared memory for slave cores. To improve
the memory bandwidth, the SW26010 processor has a 64K local cache
memory, also called local data memory (LDM), for slave cores. The
slave cores can access the data stored in the LDM with high speed.
Consequently, to improve the efficiency, one should appropriately
use the LDM to reduce the access of data in shared memory by the slave
cores and to minimize the communication between master and slave cores.
To achieve these goals, the program should be supported by the master-slave
acceleration libraries of the Sunway TaihuLight platform.

\begin{figure}
\includegraphics[scale=0.6]{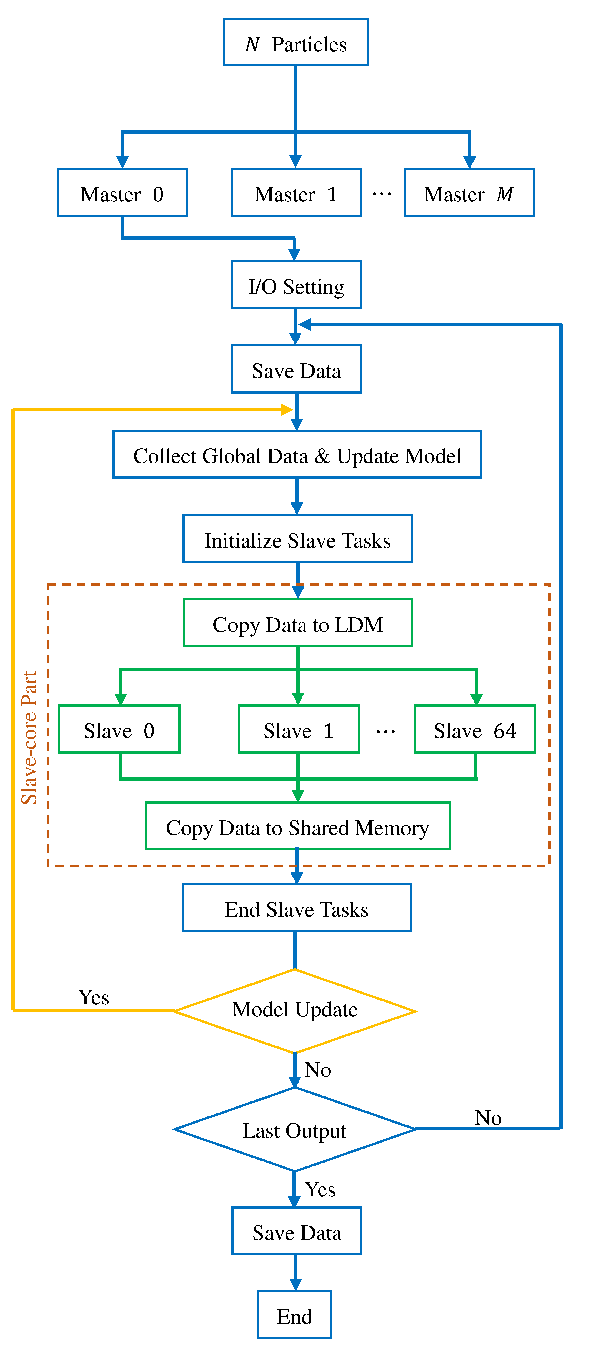}

\caption{Schematic of master-slave acceleration of APT-SW on the Sunway TaihuLight
supercomputer. The total number of particle samples is $N$, and $M$
is the number of master processes. Because the data can only be written
into the file system through master processors and the efficiency
of communication between master and slave cores is low, $n$ iterations
are divided into $m$ groups based on the users' output configuration,
and the data saving executes for $m$ times. In long-term simulations,
$n$ is much larger than $m$. Therefore, the time of master-slave
communications can be neglected compared with the iteration time.
In the brown dashed box, the slave-core code of APT-SW is depicted.
Between two processes of data saving, the master core demands each
slave core to execute iterations of $n/m$ times. The quasi-synchronous
mode is implemented to overcome the communication bottleneck during
simulations of multi-scale self-consistent problems. \label{fig:APT_SW}}
\end{figure}

As a branch version of the APT code, APT-SW has been successfully
distributed on the Sunway TaihuLight supercomputer. APT-SW remains
most of the attributes of APT while several modifications have been
made to adapt to the computation environment of the Sunway TaihuLight.
To implement the master-slave acceleration, the APT-SW source code
is divided into two parts, namely, the master-core code and the slave-core
code. As is implied by their names, the master-core code is executed
by master cores and is in charge of the data I/O, the simulation initialization,
the process parallelization, while the slave-core code is run on slave
cores, and is responsible for the particle pusher with geometric algorithms,
the electromagnetic field module, the external force module, and the
extendible module. Through the management functions, the master-core
commands can assign computation tasks to slave cores. Specifically,
the slave-core code is integrated to be one function, and the master-core
code determines when and on which slave core the slave-core function
should be executed. All the variables used in slave-core code are
independent with master-core code to avoid unnecessary access of shared
memory by slave cores. The communications between the LDM and shared
memory happens only at the stages of initialization, synchronization,
data output, and finalization of slave processes.

Figure \ref{fig:APT_SW} shows the schematic of master-slave acceleration
of APT-SW on the Sunway TaihuLight. The $N$ particle samples are
assigned to $M$ master cores, each of which manages $64$ slave cores.
One slave core executes the iteration of $N/64M$ particle samples.
Consequently, the total simulation time is mainly determined by the
computation ability and burden of each slave core. Because only the
master cores can access the file system, to record the running data,
we divide the $n$ iteration steps into $m$ groups, where $m$ is
the step number of outputs. When a master core assigns tasks to its
slave processes, the data needed are copied to the LDM. Then, the
64 slave cores push their own particles for $n/m$ steps. After that,
the data of particles will be gathered back to shared memory for output.
As a result, the amount of communications between the master and slave
core is proportional to $m$, and the time resolution of output data
is $T/m$, where $T$ is the physical time in simulations. One can
adjust the value of $m$ to find a balance between the time efficiency
and the resolution of simulation data. For secular simulations of
multi-timescale processes, the total number of iteration $n$ can
be very large. It is not necessary to save all the iteration data
to analyze the macroscopic behaviors. So the value of $m$ is much
smaller than $n$, which makes the time of master-slave communications
ignorable compared with the iteration time. For large-scale simulation
on the Sunway TaihuLight computer, the frequency of global data access
should be reduced as much as possible in order to avoid the communication
delay. As a result, the quasi-synchronous parallel calculation mode
is helpful to improve the efficiency of large-scale simulations. By
use of APT-SW on the Sunway TaihuLight supercomputer, we have successfully
completed the runaway dynamical simulations of $10^{18}$ particle-steps
which corresponds to $10^{22}$ floating-point operations. New physical
results are obtained, which will be discussed in Sec.\,\ref{sec:Taihu_results}.

\subsection{Use and extend APT\label{sub:Use-and-extend}}

The APT code aims to provide a flexible and extendible software platform
for studying and using geometric algorithms. In this section, we introduce
APT in two aspects. First, for users who just need to use the built-in
functions of the APT code, we show how to install and run the binary
file of APT, how to set the parameters via the configuration system,
and how to access the output data. Second, for researchers who would
like to study their own algorithms or physical models on APT, we provide
detailed introductions of the function interfaces, and exhibit the
three-step procedure to generate the extended source code of APT via
the extendible module.

\subsubsection{Installation }

The APT code is written in C language, and can be built by compilers
supporting the c89 standard. APT depends on the library of Lua-5.2
and Hdf5-1.8. The Lua library has been merged into the source code
of APT, while the Hdf5 library, however, should be installed before
compiling and running APT. One can visit the official website of the
Hdf5 Group to get the source code or pre-built libraries of Hdf5 software
\cite{Hdf5_web}. To install APT, users should change the working
directory to the root directory of APT source code, and simply run
the command:\begin{lstlisting}[language={bash}]
make HDF5_ROOT=<Hdf5 Directory>
make install PREFIX=<Installation Directory>
\end{lstlisting}Here, ``HDF5\_ROOT'' is the root directory of Hdf5 library, and
``PREFIX'' is the directory of installation. By deflaut, HDF5\_ROOT=/usr/local,
and PREFIX=./DataAnalysis. In the installation directory, there are
the binary file of APT called ``APT.out'', the manage configuration
file ``Config.lua'', the package of main configuration files ``pkg/'',
the document directory for configuring APT ``doc/'', and the auxiliary
tool package ``tools/''.

\subsubsection{The configuration system of APT}

The configuration system of APT is built based on the Lua language
\cite{Lua_web}. Through the Lua library, the global variables in
Lua script can be loaded by the main procedure of APT. The interface
of APT for Lua scripts supports the transfers of floating-point numbers,
integer numbers, strings, and arrays. APT provides a package of main
configuration files. Users can select a main configuration file via
the ``dofile'' command in the manage configuration file ``Config.lua''.
For example, the command \begin{lstlisting}[language={[5.2]Lua}]
dofile("./pkg/UniformField.lua");
\end{lstlisting}selects the main configuration file called ``UniformField.lua''
in ``pkg/''. Each main configuration file corresponds to one specific
task. All the main configuration file templates locate in ``pkg/'',
and their instructions can be found in the file ``doc/MainConfigFiles.txt''.
When APT is executed, the manage configuration file is first loaded,
which calls the main configuration file selected by a user. Then,
the parameters in the main configuration file will be loaded into
the parameter container of APT, which control the operation of main
procedure.

\begin{table}
\begin{tabular}{|c|c|c|c|}
\hline 
Names & Symbols & Units & Lua Global Varaible\tabularnewline
\hline 
\hline 
Mass & $m$ & $\mathrm{m_{e}}$ & Unit\_Mass\tabularnewline
\hline 
Charge & $q$ & $\mathrm{e}$ & Unit\_Charge\tabularnewline
\hline 
Time, Proper Time & $t$, $\tau$ & $\mathrm{m_{e}}/\mathrm{e}B_{0}$ & Unit\_Time\tabularnewline
\hline 
Position, Canonical Coordinate & $\mathbf{x}$, $\mathbf{X}$ & $\mathrm{m_{e}c}/\mathrm{e}B_{0}$ & Unit\_Space\tabularnewline
\hline 
Mechanical, Canonical Momentum & $\mathbf{p}$, $\mathbf{P}$ & $\mathrm{m_{e}c}$ & Unit\_P\tabularnewline
\hline 
Velocity & $\mathbf{v}$ & $\mathrm{c}$ & Unit\_V\tabularnewline
\hline 
Electric field & $\mathbf{E}$ & $B_{0}\mathrm{c}$ & Unit\_E\tabularnewline
\hline 
Magnetic field & $\mathbf{B}$ & $B_{0}$ & Unit\_B\tabularnewline
\hline 
Vecter field & $\mathbf{A}$ & $\mathrm{m_{e}}\mathrm{c}/\mathrm{e}$ & Unit\_A\tabularnewline
\hline 
Scalar field & $\phi$ & $\mathrm{m_{e}}\mathrm{c}^{2}/\mathrm{e}$ & Unit\_Phi\tabularnewline
\hline 
Hamiltonian, Energy & $\mathcal{H}$, $\mathcal{E}$ & $\mathrm{m_{e}}\mathrm{c}^{2}$ & Unit\_Energy\tabularnewline
\hline 
\end{tabular}

\caption{Units of all the physical quantities used in this paper. $\mathrm{m_{e}}$
is the rest mass of a particle, $\mathrm{e}$ is the elementary charge,
$\mathrm{c}$ is the speed of light, and $B_{0}$ is the reference
strength of magnetic field. The variables in last column can be accessed
in Lua script after the function ``GAPS\_APT\_LuaConfig\_LoadUnits($B_{0}$)''
is called.\label{tab:Units}}
\end{table}

Figure \ref{fig:TypicalConfig} shows a typical main configuration
file of APT, which is used to simulate the motion of charged particles
in uniform electromagnetic fields. The parameters are divided into
two parts, the physical parameters and the APT dimensionless parameters.
As a mandatory rule, all the variables loaded by the main procedure
of APT should be dimensionless. The units of main physical quantities
are listed in Tab.\,\ref{tab:Units}. The physical parameters would
not be loaded by APT, and are mainly used to set the physical units.
The function, ``GAPS\_APT\_LuaConfig\_LoadUnits'', on line 8 in
Fig.\,\ref{fig:TypicalConfig}, can help users calculate all the
physical units and loads them as Lua global variables listed in Tab.\,\ref{tab:Units}.
The APT dimensionless parameters presented in Fig.\,\ref{fig:TypicalConfig}
are key variables for obtaining correct simulation results. Actually,
the dimensionless parameters in one main configuration file form a
subset of the parameter container of APT. They might be used anywhere
in the main procedure. While writing a main configuration file, users
just need to ensure that all the dimensionless parameters are defined
and assigned with correct value. Then, they can organize the expressions
in main Lua configuration files in any ways they like. Because of
the flexibility of Lua script, the configuration system of APT provides
user with large freedom during parameter setting. The main procedure
of APT also automatically generates two scripts, ``LoadCalInfo.m''
and ``LoadCalInfo.py'', for matlab and python, through which users
can load all the inputted parameters during data analysis.

In Fig.\,\ref{fig:TypicalConfig}, all parameters are provided with
introductions. Most of them are easy to understand. Parameters named
with prefixes ``EMField'', ``ExtForce'', ``Pusher'', and ``Init''
are related to the configurations of the electromagnetic module, the
external force module, the particle pusher module, and the initialization
module, respectively. ``EMField\_Type'' sets the type of electromagnetic
fields. The available electromagnetic fields and their necessary setting
parameters are listed in the file ``doc/EMFields.txt''. The parameters
for enabling the calculations of external non-electromagnetic forces
are not listed in Fig.\,\ref{fig:TypicalConfig}. These variables
are named as ``ExtForce\_Cal\_<Name>''. For example, to open the
calculation of radiation force of particles, one needs to set ``ExtForce\_Cal\_RadLarmor''
as 1. The file ``doc/ExternForces.txt'' contains information for
available forces. ``Pusher\_Type'' sets the type of pushers. The
introductions for all available pushers can be found in the file ``doc/Pushers.txt''.
The methods for initializing the position and momentum of particles
are determined by ``Init\_X\_Type'' and ``Init\_P\_Type'', respectively.
For details of available initial conditions and their parameters,
one can refer to the files in directory ``doc/Initialization/''.
The other parameters also play key roles in the operation of APT.
For example, APT saves data every ``SaveSamplePeriod'' steps, and
``Open\_Work\_Cal'' can enable the calculation of works of electric
field and other non-electromagnetic forces if it is set to be 1. Another
important parameter we need to introduce is ``RunCheckPoint''. If
its value is 1, APT can execute data recovery from the point of interruption
and re-initialization automatically. This function is necessary for
large-scale simulations on supercomputers because the probability
of system exceptions increases with the scale of parallelization.
For the introductions of all parameters in APT, users can refer to
the file ``doc/Parameters.txt''.

\begin{figure}
\begin{lstlisting}[language={[5.2]Lua}]
-- Configuration File: Charged particles in uniform electromagnetic fields

-- Physical Parameters
B0 = 1; -- Magnetic strength (T)
E0 = 1; -- Electric strength (V/m)
R0 = 1.7; -- Major radius of torus (m)
a  = 0.4; -- Minor radius of torus (m)
GAPS_APT_LuaConfig_LoadUnits(B0); -- Load all units based on B0 
  
-- APT Dimensionless Parameters
num_steps = 1000; -- Simulate 1000 steps
num_total_particles = 1; -- Simulate 1 particle
OpenDataSaving = 1; -- Open data saving
SaveSamplePeriod = 1; -- Sample period for saving data
Pusher_Type = 0; -- Select pusher 0 (Volume-preserving Algorithm)
EMField_Type = 0; -- Select electromagnetic field 0 (Uniform Field)
EMFeild_Uniform_AngleEB =0; -- Angle between B and E is 0
EMField_B0 = B0/Unit_B; -- Set dimensionless magnetic strength
EMField_E0 = E0/Unit_E; -- Set dimensionless electric strength
EMField_Cal_B = 1; -- Open calculation of magnetic field
EMField_Cal_E = 1; -- Open calculation of electric field
Init_X_Type = "Constant"; -- Initial position distribution is "Constant"
Init_X_Constant_X0 = {0, 0, 0}; -- Set boundary
Init_P_Type = "Constant"; -- Initial momentum distribution is "Constant"
Init_P_Constant_P0 = {1, 0, 0.1}; -- Value of constant momentum
dT = 0.1; -- Step time length
RunCheckPoint=0; --Continue calculation from break point
\end{lstlisting}

\caption{An example of configuration files (pkg/UniformField.lua). This file
provides all the parameters for simulating charged particles moving
in a uniform electromagnetic field.\label{fig:TypicalConfig}}
\end{figure}

In current version, APT has integrated various volume-preserving and
symplectic algorithms for both classical and relativistic systems.
The volume-preserving algorithms (VPA) include both non-relativistic
and relativistic VPAs of different orders \cite{Qin_Boris_2013,Ruili_VPA_2015,HeYang_HigherOrderVPA_2016,HeYang_Spliting_2015}.
The symplectic algorithms built in APT contain the canonical symplectic
Euler algorithm, the implicit mid-point symplectic algorithm, the
K-symplectic algorithm, and the explicit symplectic algorithms based
on generating functions \cite{HeYang_Ksymp_PLA_2016,ZhangRuili_ExpGenerateSym_2016,ZhangRuili_ExpGenerateSym_Rel_2016,Geometric_numerical_integration}.
Canonical or non-canonical symplectic structures are preserved during
iteration by these symplectic algorithms. Through the preservation
of geometric structures, the geometric algorithms in the particle
pusher module equip APT with long-term stability. For example, it
has been verified that VPAs can trace the dynamics of particle correctly
for tens of billion steps \cite{Ruili_VPA_2015,HeYang_HigherOrderVPA_2016,HeYang_Spliting_2015,CollisionlessScater_NF_Letter_2016,RELong_WangYulei2016}.
Besides the geometric algorithms, traditional algorithms, such as
the Runge-Kutta algorithms of different orders, are also available
in APT. These traditional algorithms can be used as references for
the analysis of new algorithms. 

In the electromagnetic field module, several configurations have been
integrated, including the tokamak field, the terrestrial magnetic
field, the uniform field, the radially non-uniform magnetic field,
and the electric oscillator field, etc. Besides the electromagnetic
field function, the scalar and vector potentials are also available
in order to support the usages of some geometric algorithms. In current
version of APT, self-consistent field calculations for runaway electrons
and charged beams have been implemented. If setting ``EMField\_Type''
as -1, users can use discrete field data obtained from experiments
or simulations to push particles. APT can accept Hdf5 data files containing
discrete electromagnetic field data $EB=\left(\mathbf{E},\mathbf{B}\right)$
or potential data $A^{\alpha}=\left(\phi,\mathbf{A}\right)$ on 3D
cubic grids. The structure and necessary datasets of the Hdf5 file
are listed in Tab.\,\ref{tab:InputHdftstruct}. $EB$ is used by
algorithms derived from Newton equations, while $A^{\alpha}$ is needed
by symplectic algorithms. To simplify the generation of discrete field
data file, in ``tools/'', we have implemented a matlab function
``GAPS\_APT\_CreateDiscreteDataFile.m'', which can transfer a $D\times n_{x}\times n_{y}\times n_{z}\times n_{t}$
5-dimentional matlab array into an APT-readable Hdf5 file. The full
path of inputted Hdf5 file can be passed to APT via parameter ``EMField\_Discrete\_Filename''. 

\begin{table}
\begin{tabular}{|c|c|c|}
\hline 
Dataset & Dimension & Description\tabularnewline
\hline 
\hline 
/Order & $1\times1$ & Types of fields: $-1$ for $EB$, $1$ for $A^{\alpha}$\tabularnewline
\hline 
/N\_grid & $4\times1$ & Numbers of spacetime grids: $\left(n_{x},n_{y},n_{z},n_{t}\right)$\tabularnewline
\hline 
/DX & $1\times1$ & Dimensionless length of space grid\tabularnewline
\hline 
/DT & $1\times1$ & Dimensionless time step length\tabularnewline
\hline 
/OriginPoint & $3\times1$ & Origin point coordinate of field region\tabularnewline
\hline 
/Data & $n_{t}\times M$ & Discrete data of fields\tabularnewline
\hline 
\end{tabular}

\caption{Data structure of Hdf5 file for inputing discrete field data. $EB$
is a 6-dimension array containing values of electromagnetic fields,
namely $\left(\mathbf{E},\mathbf{B}\right)$, and $A^{\alpha}$ is
a 4-dimensional array containing scalar and vector potentials, namely
$\left(\phi,\mathbf{A}\right)$. $n_{x}$, $n_{y}$ and $n_{z}$ denote
numbers of spatial grids on each dicrection, and $n_{t}$ is the number
of time steps of discrete field data. For static fields, one should
set $n_{t}=1$. $x_{0}$, $y_{0}$, and $z_{0}$ are dimensionless
coordinates of the origin point of field region. For each moment,
the data of a tensor $F_{ijk}^{l}$ should be rewritten as a 1D array
$F_{J}$ based on the index mapping $J=l+D\left[i+n_{x}\left(j+n_{y}k\right)\right]$,
where $D$ is the number of components of $\mathbf{F}$, $i$, $j$,
$k$ are indexes of grids on three directions, and $l$ is the index
of components of $\mathbf{F}$. $M=D\times n_{x}\times n_{y}\times n_{z}$
is the number of data for each moment. The value of $D$ can be set
as 6 or 4 for $EB$ or $A^{\alpha}$, respectively. \label{tab:InputHdftstruct}}
\end{table}

The external force-field module loads non-electromagnetic forces into
the dynamical simulation of particles. Various external forces including
synchrotron radiation force, collisional force, bremsstrahlung force,
and gravitation field have been built in the APT code. To calculate
the random collisional force, the Langevin Coulomb collision model
is adopted, which is an energy-preserving random numerical method
for plasmas without electric field \cite{Cadjan_LangevinCoulombCollision_1999}.

Through the parameters ``Init\_X\_Type'' and ``Init\_P\_Type'',
one can set different types of initial distributions. The initialization
module provides several commonly used methods of statistical sampling.
For position sampling, APT supports uniform distributions of different
shapes, including cuboid, cylinder, and torus. For momentum sampling,
APT supports two-stream sampling, normal-distribution sampling, etc.
The initialization module facilitates the usage of APT in complex
realistic physical problems.

\subsubsection{The architecture of Hdf5 output data}

The Hdf5 library provides a file-system-like way of reading and writing
data \cite{Hdf5_web}, i.e., one can access the data like manipulating
a file in Linux operation system. This mechanism can simplify the
procedure of accessing and analyzing data. The Hdf5 format is supported
by most of the popular data-analysis tools, such as Matlab and Python. 

APT stores simulation data in an Hdf5 file called ``Data.h5''. The
basic dataset is ``/PTC'', which contains all the information of
particles. ``/PTC'' is a $L_{s}N_{ptc}\times N_{steps}$ 2-dimensional
matrix. Here $N_{ptc}$ and $N_{steps}$ are respectively the number
of particles and the number of saved steps. $L_{s}=11$ is the number
of data for one particle. The sequence of data for one particle is
$S_{die}$, $t$, $x$, $y$, $z$, $p_{x}$, $p_{y}$, $p_{z}$,
$a_{x}$, $a_{y}$, $a_{z}$. If the value of $S_{die}$ is 1, the
particle is ``dead'' and the program stops updating its data. The
main procedure of APT changes the value of $S_{die}$ to 1 if a particle
is out of boundary of discrete fields. The other data are time ($t$),
Cartesian coordinates ($x,y,z$), momentum/velocity ($p_{x},p_{y},p_{z}$),
and acceleration ($a_{x},a_{y},a_{z}$). All of them are dimensionless.

If enabling the calculations of electromagnetic fields or other external
fields, users can also find datasets in ``Data.h5'' named after
these forces in the Hdf5 file. They are all $3\times N_{steps}$ matrices.
Moreover, if the work calculation is enabled, the data of works for
all forces, except magnetic force, will be stored in datasets named
as ``/Work\_<Name>'', where, <Name> is the name of forces. These
work datasets are ``$1\times N_{steps}$'' matrices.

\subsubsection{The interfaces of APT}

Once being familiar with the installation procedures, the configuration
system, and the data format, users can benefit from all the built-in
resources of algorithms and physical modules of APT. However, they
can do more than that. In the field of scientific calculation, researchers
usually need to extend codes to achieve their personal goals. Sometimes,
they need to understand the architecture of a code and modify different
places of the source code. This process takes researchers a lot of
time and reduces the efficiency for researching. And, because there
is no standard rule for modification and extension, most of the modified
codes cannot be used easily by other researchers. APT aims to improve
the efficiency of extending the source code, integrate novel works
from different fields, and finally provide researchers a platform
of cooperation on which the geometric algorithms can be studied and
used more conveniently. To achieve this goal, the APT project designs
standard function interfaces and naming rules for the containers of
variables, pushers, electromagnetic fields, external forces, and initialization
methods. Before showing how to extend APT, we first give a detailed
description about the interfaces of APT in this part.

\paragraph{The parameter container}

As we have introduced before, the operation of APT is governed by
lots of control parameters loaded from configuration files. Therefore,
the parameter container is a vital part of APT, which appears in most
of interfaces. The parameter container is a structure named ``Gaps\_IO\_In\-puts\-Container'',
whose fields compose the complete set of input parameters. Its definition
is:

\begin{lstlisting}[language={C}]
typedef struct{ 
     long num_steps;
     double EMField_B0;
     char Init_X_Type[50];
     double Init_P_Constant_P0[3];
//   ...
}Gaps_IO_InputsContainer;
\end{lstlisting}Through this structure, developers can access all the control parameters.

\paragraph{The particle structure}

Like the parameter container structure, the particle structure is
also an important data type for interfaces of APT, which is named
as ``Gaps\_APT\_Particle''. The particle structure contains all
the data of a particle, such as time, position, momentum, acceleration,
and electromagnetic field. Through the particle structure, developers
can access all the particle data and functions of electromagnetic
fields and external forces. APT provides a set of functions for obtaining
data in a particle structure. These functions take the pointer of
a particle structure as its input, and return the pointers of corresponding
data. They are named following the format ``GAPS\_A\-PT\_G\-et<Name><Dim>'',
where <Name> is the abbreviation of a physical quantity, and <Dim>
is a integer denoting the dimension of data. Here is a typical example
for obtaining the pointer of 4-dimensional space time:

\begin{lstlisting}[language={C}]
double * GAPS_APT_GetX4(Gaps_APT_Particle *pPtc);     // SpaceTime
\end{lstlisting}The complete list of these functions is given in Tab.\,\ref{tab:ptcstructData}.

\begin{table}
\begin{tabular}{|c|c|}
\hline 
Name & Physical quantity\tabularnewline
\hline 
\hline 
GAPS\_APT\_GetDie1 & Marker of life\tabularnewline
\hline 
GAPS\_APT\_GetCharge1 & Charge\tabularnewline
\hline 
GAPS\_APT\_GetMass1 & Mass\tabularnewline
\hline 
GAPS\_APT\_GetS1 & Proper time\tabularnewline
\hline 
GAPS\_APT\_GetX4 & Space time\tabularnewline
\hline 
GAPS\_APT\_GetP4 & 4-momentum\tabularnewline
\hline 
GAPS\_APT\_GetCanP4 & 4-canonical-momentum\tabularnewline
\hline 
GAPS\_APT\_GetA4 & 4-vector-potential\tabularnewline
\hline 
GAPS\_APT\_GetT1 & Time\tabularnewline
\hline 
GAPS\_APT\_GetX3 & Position\tabularnewline
\hline 
GAPS\_APT\_GetP3 & Momentum\tabularnewline
\hline 
GAPS\_APT\_GetCanP3 & Canonical momentum\tabularnewline
\hline 
GAPS\_APT\_GetA3 & Vector potential\tabularnewline
\hline 
GAPS\_APT\_GetGamma1 & Lorentz factor\tabularnewline
\hline 
GAPS\_APT\_GetE3 & Electric field\tabularnewline
\hline 
GAPS\_APT\_GetB3 & Magnetic field\tabularnewline
\hline 
GAPS\_APT\_GetAclr3 & Acceleration\tabularnewline
\hline 
GAPS\_APT\_GetEwork1 & Work of electric field\tabularnewline
\hline 
GAPS\_APT\_GetFieldFunc & Function pointer of EM field\tabularnewline
\hline 
\end{tabular}

\caption{List of functions for accessing data from the particle structure.\label{tab:ptcstructData}}
\end{table}

\paragraph{The electromagnetic field interface}

Geometric algorithms usually have complex forms. Sometimes high-order
derivatives and integral calculation of field functions are needed.
The electromagnetic field module aims to provide a unified interface
for calculating not only electromagnetic field functions, but also
their integrals and derivatives. Therefore, the prototype of field
functions is defined as:\begin{lstlisting}[language={C}]
typedef int (*Gaps_APT_Field)(double *pTensor,double *pSpaceTime4,int Order,
                              Gaps_IO_InputsContainer *pInputs);
\end{lstlisting}

\begin{table}
\begin{tabular}{|c|c|c|c|}
\hline 
Tensor & Order & Dimension & Length of data array\tabularnewline
\hline 
\hline 
$EB$ & $-1$ & $6$ & $6$\tabularnewline
\hline 
$A^{\alpha}$ & $1$ & $4$ & $4$\tabularnewline
\hline 
$\partial_{\beta}A^{\alpha}$ & $2$ & $4$ & $16$\tabularnewline
\hline 
$\partial_{\beta}\partial_{\gamma}A^{\alpha}$ & $3$ & $4$ & $64$\tabularnewline
\hline 
$\cdots$ & $\cdots$ & $\cdots$ & $\cdots$\tabularnewline
\hline 
\end{tabular}

\caption{Definitions of tensors in APT. $A^{\alpha}=\left(\phi,\mathbf{A}\right)$
is the 4-potential vector, and $\partial_{\alpha}$ denotes the derivative
operation with respect to the $\alpha$-th component of spacetime.
Definitions of tensors corresponding to $A^{\alpha}$ are the same
with ordinary conventions. The electromagnetic tensor $F_{\alpha\beta}=\partial_{\alpha}A_{\beta}-\partial_{\beta}A_{\alpha}$
is transformed to a 6-dimensional 1-order array $EB=\left(\mathbf{E},\mathbf{B}\right)$
in APT to avoid the confusion of the ``4-dimensional 2-order'' tensor.\label{tab:tensordef}}
\end{table}

The function interface uses the 4-dimensional representation, which
takes 4-dimensional spacetime, $X^{\alpha}=\left(t,\mathbf{X}\right)$,
as input parameters and returns tensors of different orders. Other
necessary inputs are the pointer of parameter container and the order
of the tensor. The definitions of tensors in APT are different from
ordinary conventions, which are listed in Tab.\,\ref{tab:tensordef}.
Tensors of 3-order can be calculated by most of electromagnetic field
functions to support the use of some high order symplectic algorithms.
The template for electromagnetic field functions is\begin{lstlisting}[language={C}]
#include "APT_AllHeaders.h"
int GAPS_APT_Field_<Name>(double *pTensor,double *pSpaceTime4,int Order,
                          Gaps_IO_InputsContainer *pInputs)
{	
	/**Step 1: Set the maximum tensor order**/
	int MaxOrder = 1;
	/**Step 2: Declare and calculate intermediate variables**/
	/**Step 3: Calculate EB: Order=-1**/
	if(-1 == Order)
	{
		if(pInputs->EMField_Cal_E)
		{
			//Calculate E and assign values to pTensor[0]~pTensor[2]
		}	
		if(pInputs->EMField_Cal_B)
		{
			//Calculate B and assign values to pTensor[3]~pTensor[5]
		}		
	}
	/**Step 4: Calculate 4-vector-potential: Order=1**/
	if(1 == Order)
	{
		//Calculate A and assign values to pTensor[0]~pTensor[3]
	}
	/**Step 5: Error detection**/
	if(MaxOrder < Order || 0 == Order)
	{
		fprintf(stderr,"ERROR: In function GAPS_APT_Field_<Name>. 
                This field function does NOT support tensor of 
				0-order or order larger than %d.\n",MaxOrder);
	}
	return 0; 
} 
\end{lstlisting}

The formal parameter ``pTensor'' in field function interface points
to the head of a 1-dimensional array, which contains the data of tensors.
The length of ``pTensor'' is determined by both the dimension and
the order of a tensor. For a $n$-order $m$-dimensional tensor, the
length of ``pTensor'' is $L_{fld}=m^{n}$, see Tab.\,\ref{tab:tensordef}.
APT has implemented functions that can return the components of ``pTensor''
through its index array, namely,\begin{lstlisting}[language={C}]
double GAPS_APT_TensorValue(double *pTensor,int *pIndex,int Dim,int Order);
double *GAPS_APT_TensorPointer(double *pTensor,int *pIndex,int Dim,int Order);
\end{lstlisting}``pIndex'' is an integer array with $n$ elements, which contains
the indexes of tensor. The ``Dim'' and ``Order'' are the dimension
and the order of a tensor, respectively.

Some geometric algorithms need to calculate the definite integral
of field functions \cite{HeYang_Ksymp_PLA_2016}. APT provides a numerical
quadrature function to accomplish the definite integral of arbitrary
components of arbitrary field tensor functions, namely,

\begin{lstlisting}[language={C}]
double GAPS_APT_FieldIntegral(int n, Gaps_APT_Particle *pPtc,
                              int Order, int *pIndex, int idxInt,
                              double *pST0,double *pST1,                              
                              Gaps_IO_InputsContainer *pInputs);
\end{lstlisting}Here, ``n'' defines the number of intervals within the integral
domain, ``Order'' is the order of tenors, ``pIndex'' is an array
denoting the component the integrand tensor, ``idxInt'' sets which
component of spacetime is the integral variable, and ``pST0'' and
``pST1'' are the start and end points of integral domain. For example,
to calculate the integral $\int_{t_{0}}^{t_{1}}EB_{1}dt=\int_{t_{0}}^{t_{1}}E_{y}dt$
for $\mathbf{X}=\left(x_{0},y_{0},z_{0}\right)$, the corresponding
codes when implementing a new pusher are \begin{lstlisting}[language={C}]
int Order=-1;//Order of EB is -1
int pIndex[1]={1};//Integrand is Ey
int idxInt=0;//The 0th component of spacetime is t.
int n=512;
double pST0[4]={a,x0,y0,z0};
double pST1[4]={b,x0,y0,z0};
double result;
result=GAPS_APT_IntegralField(n,pPtc,Order,pIndex,idxInt,pST0,pST1,pInputs);
\end{lstlisting}Here, ``pPtc'' points to a particle structure, ``pInputs'' points
to the parameter container.

APT approximates the self-consistent field of particles through mapping
the phase-space states of all particles to several dynamical parameters
of field functions. Generally, for any self-consistent model, one
needs the information of all particles at a moment. Therefore, we
provide a function to gather the information of all particles, namely,\begin{lstlisting}[language={C}]
int GAPS_APT_GatherPtcInfo(double *pData);
\end{lstlisting}This function returns the positions and momenta of all particles into
a $6N$-dimensional array ``pData''. The data sequence in ``pData''
is $\mathbf{x}_{0}$, $\mathbf{p}_{0}$, $\mathbf{x}_{1}$, $\mathbf{p}_{1}$, $\cdots$, $\mathbf{x}_{i-1}$, $\mathbf{p}_{i-1}$.
Through this function, user can accomplish different approximate self-consistent
models of particle dynamics. This simplified self-consistent model
of APT is designed to find a compromise between efficiency and accuracy
when the self-consistent calculation cannot be avoided. For many specific
physical processes, this method can still provide useful physical
information with high efficiency.

Finally, for discrete field data, the field function interface is
the same with analytical functions. The local fields of a particle
are calculated through the interpolation of discrete data. Users need
not to know the details of discrete field structure, but can use the
field function of discrete field data just like the analytical field
functions.

\paragraph{The external non-electromagnetic force interface}

The prototype of non-electro\-magnetic force functions is: \begin{lstlisting}[language={C}]
typedef int (*Gaps_APT_ExtForce)(double *pForce3,Gaps_APT_Particle *pPtc,
                                 Gaps_IO_InputsContainer *pInputs);
\end{lstlisting}The external force module works with the algorithms derived from Newton
equations. The dimension of force vector is limited to be 3. The pointer
of particle structure is passed into the force functions to provide
all needed data. The constant coefficients of a force can be given
through the parameter container. The template of external non-elec\-tromag\-netic
force functions is\begin{lstlisting}[language={C}]
#include "APT_AllHeaders.h"
int GAPS_APT_ExtForce_<Name>(double *pForce3,Gaps_APT_Particle *pPtc,
                            Gaps_IO_InputsContainer *pInputs)
{	
	/**Step 1: Get pointers of particle data**/
	double *pX = GAPS_APT_GetX3(pPtc);
	/**Step 2: Get constant coefficients from pInputs**/
	/**Step 3: Calculate and assign values to pForce3[0]~pForce3[2]**/
	return 0; 
} 
\end{lstlisting}

In APT, each external force is assigned with a serial integer number.
Developers can calculate the value of a force through function\begin{lstlisting}[language={C}]
int GAPS_APT_CalExtForce(double *pForce3,int Type,Gaps_APT_Particle *pPtc,
                         Gaps_IO_InputsContainer *pInputs);
\end{lstlisting}This function return the value of the force with serial number ``Type''.
To update all the force data in particle structure, the function \begin{lstlisting}[language={C}]
int GAPS_APT_UpdatePtcData_ExtForce(Gaps_APT_Particle *pPtc,
                                    Gaps_IO_InputsContainer *pInputs);
\end{lstlisting}is useful. If users want to obtain the sum of all the loaded forces
and update the force data at the same time, they can call\begin{lstlisting}[language={C}]
int GAPS_APT_MergeExtForce(double *F_ext3,Gaps_APT_Particle *pPtc,
                           Gaps_IO_InputsContainer *pInputs);
\end{lstlisting}These three functions are useful during implementing a new particle
pusher.

\paragraph{The particle pusher interface}

The particle pusher interface is the most important part of APT. The
prototype of particle pusher functions is defined as\begin{lstlisting}[language={C}]
typedef int (*Gaps_APT_ParticlePusher)(Gaps_APT_Particle *pPtc,
                                        Gaps_IO_InputsContainer *pInputs);
\end{lstlisting}When writing a pusher function, one should ensure the pusher can finish
two tasks: 1) for all pushers, update the time, position, and the
mechanical momentum of a particle through an algorithm, 2) for pushers
derived from Newton equations, update electromagnetic data and non-electromagnetic
force data. Here is a template for a particle pusher\begin{lstlisting}[language={C}]
#include "APT_AllHeaders.h"
int GAPS_APT_Pusher_<Name>(Gaps_APT_Particle *pPtc,
                         Gaps_IO_InputsContainer *pInputs)
{
	/**Step 1: Get pointers of particle data**/
	double *pT = GAPS_APT_GetT1(pPtc);//Time
	double *pX = GAPS_APT_GetX3(pPtc);//Position	
	double *pP = GAPS_APT_GetP3(pPtc);//Momentum
	double *E = GAPS_APT_GetE3(pPtc);//Electric field
	double *B = GAPS_APT_GetB3(pPtc);//Magnetic field
	Gaps_APT_Field FieldFunc= GAPS_APT_GetFieldFunc(pPtc);
	/**Step 2: Get parameters from pInputs**/
	double dT=pInputs->dT;//Time step length	
	/**Step 3: Update pT, pX, pP through an algorithm**/ 
	/**Step 4: Update electromagnetic field data**/
    GAPS_APT_UpdatePtcData_EB(pPtc,pInputs);
	/**Step 5: Update force data**/	
    GAPS_APT_UpdatePtcData_ExtForce(pPtc,pInputs);
	return 0; 
} 
\end{lstlisting}While implementing a new pusher function, one needs not to know the
specific forms of electromagnetic fields or external forces, which
have been built in other modules. They just need to focus on their
own job of algorithms.

\paragraph{The initialization interface}

Researcher can add different approaches of initialization into APT
through initialization interface. The interfaces for position and
momentum initialization are \begin{lstlisting}[language={C}]
int GAPS_APT_SetParticlePosition_<Name>(Gaps_APT_Particle *pPtc,
                                        Gaps_IO_InputsContainer *pInputs);
int GAPS_APT_SetParticleMomentum_<Name>(Gaps_APT_Particle *pPtc,
                                        Gaps_IO_InputsContainer *pInputs);
\end{lstlisting}``<Name>'' denotes the name of initialization methods. For example,
the function for Maxwell distribution of momentum is ``GAPS\_APT\_Set\-Particle\-Momentum\_Ma\-xwell'',
which can be loaded by APT if the parameter ``Ini\-t\_P\_T\-ype'' in
configuration file is set as ``Maxwell''. The control parameters
of generating distributions can be accessed via ``pInputs''. The
template of initialization function is \begin{lstlisting}[language={C}]
#include "APT_AllHeaders.h"
int GAPS_APT_SetParticlePosition_<Type>(Gaps_APT_Particle *pPtc,
                                        Gaps_IO_InputsContainer *pInputs)
{
	/**Step 1: Get pointers of coordinates**/
	double *pX = GAPS_APT_GetX3(pPtc);
	/**Step 2: Get parameters from pInputs**/
	/**Step 3: Generate random number and assign them to pX*/
	return 0; 
} 
\end{lstlisting}

\subsubsection{Extend APT\label{sub:APT_Exten}}

In this part, we give a detailed introduction on how to extend the
configuration parameters, the electromagnetic functions, the external
forces, the particle pushers, and the initialization approaches of
APT.

\paragraph{A brief description of the extendible module}

The extendible module of APT is implemented based on Bash-script.
Following three steps, users can conveniently build a new version
of APT for new applications, see Fig.\,\ref{fig:APT_Arch}. First,
write new C functions for electromagnetic fields, external forces,
algorithms, or initialization approaches in specified directories
of APT. Second, modify the file ``script/ADD'' and tell APT the
names, types and introductions of these new functions, and the necessary
parameters. Third, run the script named ``Generation.sh''. Then,
all the source codes and documents will be generated automatically.
The extended version of APT can be compiled directly. After writing
a Lua configuration file, users can test and use their own functions
immediately. The detailed three-step procedures are listed as follows.

\paragraph{Step 1: Write new function files}

When introducing the interfaces of APT, we have provided templates
for electromagnetic field functions, external force functions, particle
pusher functions, and initialization functions. Users need to organize
one function in a single file. Then they should move new function
files into corresponding directories. The electromagnetic function
files locate at ``src/EMField/'', the external force function files
locate at ``src/ExtForce'', the particle pusher function files are
placed at ``src/Pusher'', and the initialization function files
are in ``src/Initialization''.

\paragraph{Step 2: Add function descriptions and new parameters}

Now developers should add descriptions for new functions via the management
script of the extensible module, ``script/ADD''. There are several
commands provided by APT to conveniently accomplish this job, namely,
\begin{lstlisting}[language={bash}]
Add_EMField    "<Name>"    "<Parameters>"    "<Notes>"    "<Info>"
Add_ExtForce   "<Name>"    "<Parameters>"    "<Notes>"    "<Info>"
Add_Pusher     "<Name>"    "<Parameters>"    "<Notes>"    "<Info>"
Add_InitMethod "<Class>"   "<Name>"    "<Parameters>"  "<Notes>"  "<Info>"
\end{lstlisting}For example, to add an electromagnetic function called ``GAPS\_APT\_F\-ield\_Un\-iform'',
the full command is\begin{lstlisting}[language={bash}]
Add_EMField "Uniform" "EMFeild_Uniform_AngleEB" "MaxOrder:3" "Uniform EM field"
\end{lstlisting}Developers should notice that <Name> here dose not denote the full
name of functions but the name without prefix. <Parameters> is a list
of all necessary parameters used in new functions, and <Info> is a
short introduction about the function. <Notes> should contain important
information to help users use the function correctly. The first input
<Class> of ``Add\_InitMethod'' can be ``X'' or ``P'', which
denote the initialization functions for positions or momenta, respectively.

For new functions, additional parameters might be used. For example,
the parameter for setting the temperature of Maxwell distribution
is ``Init\_P\_Maxw\-ell\_Temp''. APT requires that all the extended
parameters are named following the standard of APT. Parameters for
electromagnetic field, external non-electromagnetic forces, particle
pushers, and initialization approaches should be named as ``EMFiel\-d\_<Nam\-e>\_<Para\-meter>'',
``ExtForce\_<N\-ame>\_<Para\-meter>'', ``Pusher\_<Na\-me>\_<Par\-ameter>'',
and ``Init\_<X/P>\_<Na\-me>\_<P\-ar\-ameter>'', respectively, where <Name>
denotes the name of functions, and <Parameter> can be designed freely
by users. The command for adding new parameters in the file ``script/ADD''
is\begin{lstlisting}[language={bash}]
Add_Inputs    "<Type>"    "<Dim>"    "<Name>"    "<Introduction>"
\end{lstlisting}<Type> is the type of parameters, which can be ``double'', ``long'',
``char''. <Dim> is a integer larger than 0 and denotes the dimension
of parameters. <Name> is the name of parameters. <Introduction> is
a short description of a parameter, which can help users understand
the functions of parameters. If <Dim> is 1, the extendible module
will declare the parameter as a single variable; if <Dim> is larger
than 1, the parameter is declared as an array.

\paragraph{Step 3: Update source code via one command}

After modifying the file ``script/ADD'', there is only one last
step needed to update the source code, namely, changing directory
to ``script/'' and run the command \begin{lstlisting}[language={bash}]
./Generate.sh
\end{lstlisting}Then all the new functions will be integrated into APT, and the documents
in ``doc/'' will also be updated at the same time. The new version
of APT source code can be directly compiled. User can also optionally
provide new main configuration files in ``pkg/''. The introductions
of them should be written in ``doc/MainConfigFiles.txt''. 

By use of the extendible module, researchers from different fields
can conveniently focus on their own tasks and use the components provided
by others directly. For example, researchers from the plasma physics,
accelerator physics, space science, and fusion energy can add the
customized electromagnetic configurations and directly use the advanced
geometric algorithms provided by computational mathematicians. The
complex physical field configurations can be used by mathematicians
to analyze their newly developed algorithms. This extendible module
can thus boost the integration of new results from different fields.

\section{Applications of APT in Scientific research\label{sec:Benchmarks}}

The APT code has been used in the studies of new algorithms and the
simulations of a variety of important plasma processes. Due to the
secular stability provided by the geometric algorithm kernel, APT
can solve many multi-scale and nonlinear problems that cannot be addressed
by traditional simulation methods, which stimulates the discoveries
of new physical phenomena. In this section, we exhibit the prospects
of APT in two applications, namely, the secular dynamics of runaway
electron in tokamak and the distribution evolution of energetic particles
in Van Allen belts.

\subsection{The Secular Dynamics of Runaway Electrons}

Runaway electrons are energetic charged particles generated in tokamaks
which are the most prospective devices for controlled fusion energy
\cite{Drercer_REorigins_1959,Connor_Relativistic_RE_1975}. During
the operation of tokamaks, fast shutdown, disruptions, and strong
current drive can induce the generation of large amounts of runaway
electrons \cite{Yoshino_Shutdown_RE,Jaspers_Disruption_RE,Helander_avalanch_2000,Helander_2002,Fulop_magneticThreshold4RE2009,Gill_REref1_2000,Jaspers_RErefs4_1993,Nygren_RErefs5_1997,Parks_RErefs6_1999,Rosenbluth_RErefs7_1997,Yoshino_REandTurbulenceDischarge2000,Tamai_Yoshino_REtermination_JT60U_2002,Lehnen_RMP_PRL_TEXTOR,Finken_RElosses_2007,Net_Fisch_RevModP1987}.
Due to the acceleration of the induced loop electric field, runaway
electrons carrying energies of tens of MeVs have been reported in
various experiments \cite{Bartels_RE_PFCs1994,Kawamura_PFS_RE1989,Bolt_REref0_1987,Jaspers_REref3_2001}.
The existence of runaway electrons is a potential threat to safety
operation of tokamak devices. To study the runaway process, the relativistic
effect and the synchrotron radiation cannot be neglected. The runaways
reach the synchrotron energy limit when the acceleration of the electric
field is balanced out by the synchrotron radiation dissipation \cite{Guan_Qin_Sympletic_RE,Martin_Momentum_RE_1998,Martin_Energylimit_RE_1999,LiuJian_RE_Positron_2014}.
Both the detailed dynamical behaviors and the rules of energy limit
of runaway electrons are important topics in the field of fusion energy.

The runaway dynamics is a typical multi-scale process, which involves
timescales from characteristic time of Lorentz force ($10^{-11}\,\mathrm{s}$)
to energy balance time ($1\,\mathrm{s}$) \cite{RELong_WangYulei2016,LiuJian_RE_Positron_2014}.
The runaway dynamics thus spans about 11 orders of magnitude in timescale.
To eliminate the trouble of multi-timescale, traditional methods average
out the motion in small timescale, which simplifies the problem but
loses physical information. However, with the secular stability of
geometric algorithms, APT can simulate directly the full-orbit dynamics
which needs more than $10^{12}$ time steps. Correspondingly, runaway
dynamical pictures containing information of all timescales have been
produced. Especially, the discovery of neoclassical collisionless
pitch-angle scattering brings a novel understanding about runaway
behaviors and results in new laws of energy limit.

\begin{figure}
\includegraphics{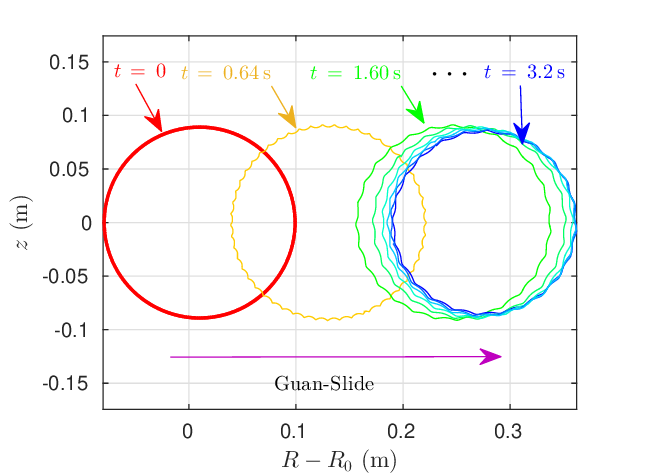}

\caption{Full-orbit snapshots of runaway orbit projected in the poloidal plane
at different moments. The configuration of field is determined by
$R_{0}=1.7\,\mathrm{m}$, $a=0.4\,\mathrm{m}$, $q=2$, $B_{0}=2\,\mathrm{T}$,
and $E_{l}=0.2\,\mathrm{V/m}$. The runaway electron is initially
sampled with momentum $p_{\parallel0}=5\,\mathrm{m_{0}c}$, $p_{\perp0}=1\,\mathrm{m_{0}c}$
at $R=1.8\,\mathrm{m}$, $\xi=z=0$. Besides the neoclassical radial
drift, the ripple structures are obviously exhibited.\label{fig:Drift}}
\end{figure}

Figure \ref{fig:Drift} depicts the snapshots of runaway orbit projected
on poloidal plane at different moments simulated by APT. Besides the
Guan-Slide effect, this full-orbit result also exhibits clearly the
detailed ripple structures in small timescale \cite{Guan_Qin_Sympletic_RE,RELong_WangYulei2016}.
To obtain Fig.\,\ref{fig:Drift}, we use the relativistic volume-preserving
algorithm in APT. Because the runaway dynamics involves the synchrotron
radiation, the external force-field module of APT is also used. The
tokamak field configuration file of APT provides the needed field
given by
\begin{equation}
\mathbf{B}=-\frac{B_{0}R_{0}}{R}\mathbf{e}_{\xi}-\frac{B_{0}\sqrt{\left(R-R_{0}\right)^{2}+z^{2}}}{qR}\mathbf{e}_{\theta}\,,\label{eq:B}
\end{equation}
\begin{equation}
\mathbf{E}=E_{l}\frac{R_{0}}{R}\mathbf{e}_{\xi}\,,\label{eq:E}
\end{equation}
where, $\mathbf{e}_{\xi}$ and $\mathbf{e}_{\theta}$ are respectively
the toroidal and poloidal unit vectors, $R_{0}$ is the major radius,
$q$ denotes safety factor, $E_{l}$ is the strength of loop electric
field, and $B_{0}$ is the magnitude of background magnetic field.
In calculation, we set parameters based on a typical tokamak, that
is $R_{0}=1.7\,\mathrm{m}$, $a=0.4\,\mathrm{m}$, $q=2$, $B_{0}=2\,\mathrm{T}$,
and $E_{l}=0.2\,\mathrm{V/m}$. The initial position of a runaway
eletron is chosen as $R=1.8\,\mathrm{m}$, $\xi=z=0$, and the initial
parallel and perpendicular momenta are set as $p_{\parallel0}=5\,\mathrm{m_{0}c}$
and $p_{\perp0}=1\,\mathrm{m_{0}c}$ respectively. The time step of
simulation is set as $\Delta t=1.9\times10^{-12}s$. The total number
of iteration steps is thus about $1.6\times10^{12}$. The long-term
stability of APT ensures the correctness of simulation results even
after $10^{12}$ steps. 

\subsection{Energetic Particles in Van Allen Radiation Belt}

Since its discovery in 1958, the Van Allen radiation belts have been
an important topic of geophysics \cite{VanAllen_RadiationBelt}. The
Van Allen belt is a region about several $R_{E}$s away from the surface
of earth, where $R_{E}$ is the radius of earth at the equator. Large
amounts of particles carrying energies of several MeVs are confined
in the radiation belt. These energetic particles do harm to the spacecraft
as well as the satellites. The mechanism of acceleration of energetic
particles in radiation belts is still not clear, even though many
works have been done on this problem \cite{VanAllen1,VanAllen2}. 

APT has offered an outstanding platform for studying the statistical
behaviors and the acceleration mechanisms of radiation-belt particles.
The background terrestrial field is approximated by 
\[
\mathbf{B}\left(\mathbf{x}\right)=\frac{B_{0}R_{0}^{3}z}{\left(R^{2}+z^{2}\right)^{2}}\mathbf{e}_{R}+\frac{B_{0}R_{0}^{3}}{R^{2}+z^{2}}\left(-\frac{R}{R^{2}+z^{2}}+\frac{1}{2R}\right)\mathbf{e}_{z}\,.
\]
The corresponding vector field is 
\begin{equation}
\mathbf{A}\left(\mathbf{x}\right)=\frac{B_{0}R_{0}^{3}}{2\left(R^{2}+z^{2}\right)}\mathbf{e}_{\theta}\,,\label{eq:A_belts}
\end{equation}
where $R_{0}=6.6R_{E}$ and $B_{0}=2B_{surf}/6.6^{3}$, $R_{E}=6370000\,\mathrm{m}$
is the radius of earth, and $B_{surf}=3.12\times10^{-5}\,\mathrm{T}$
is the magnetic field on surface of earth. The parallelization module
is used to achieve large-scale statistical simulations on supercomputers.
Users can choose different geometric algorithms, adopt various initial
statistical samplings, and test different electromagnetic-wave acceleration
processes in APT.

\begin{figure}
\includegraphics[scale=0.7]{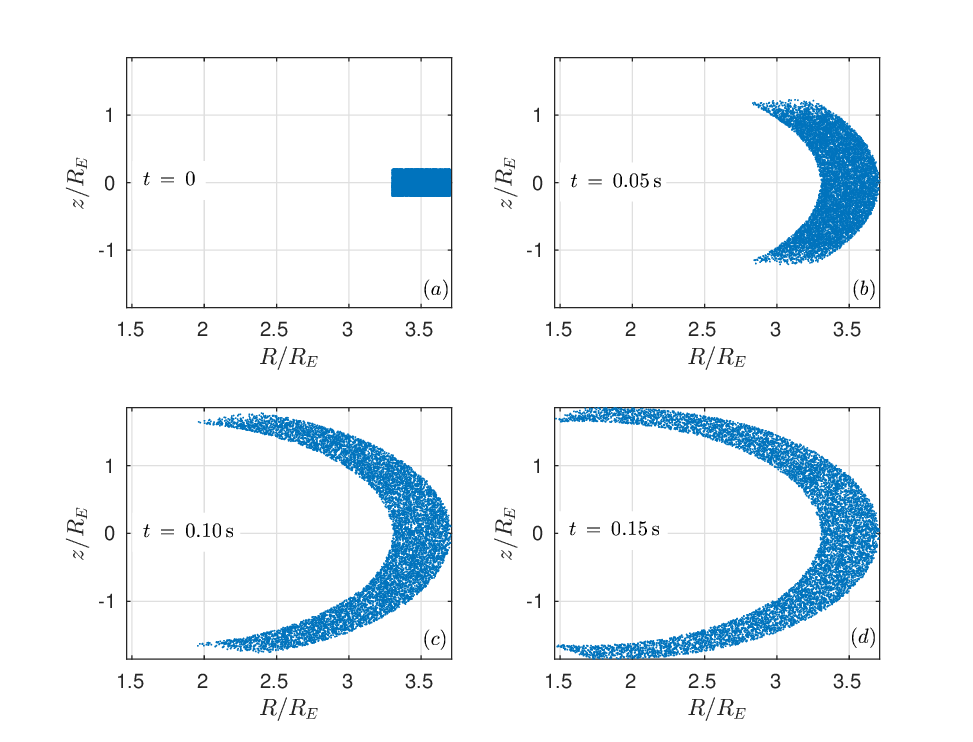}

\caption{The diffusion process of energetic electrons in Van Allen belts.\label{fig:VanAllen}}
\end{figure}

We choose the symplectic Euler algorithm for relativistic charged
particles to simulate the collective evolution of energetic electrons
in the background magnetic field of earth. Initially, $10^{4}$ particles
are uniformly sampled in the region $3.3R_{E}\leqslant R\leqslant3.7R_{E}$
and $-0.2R_{E}\leqslant z\leqslant0.2R_{E}$. The initial Lorentz
factors of particles are given by a normal distribution with the average
value $\mu=0.6\,\mathrm{MeV}$ and the standard deviation $\sigma=0.5\,\mathrm{KeV}$,
and the initial momenta are sampled uniformly within $-1\leqslant p_{\parallel}/p\leqslant1$.
The time step is about $5.24\times10^{-6}\,\mathrm{s}$, and the number
of steps is $2.8\times10^{5}$. The diffusion process of energetic
electrons in Van Allen belt of earth is shown in Fig.\,\ref{fig:VanAllen}.
Since the confined particles in the initial region have different
momenta and phases, they do not move together but spread in the earth
magnetic field. The profile of distribution at $t=0.15\,\mathrm{s}$
shows the shape of radiation belts, which reveals the formation of
Van Allen belt. 

\section{Large-scale Simulation of Runaway electrons on the Sunway TaihuLight
Supercomputer\label{sec:Taihu_results}}

Based on APT-SW, the long-term evolution of runaway beam is simulated
under parameters of ITER tokamak on the Sunway TaihuLight supercomputer.
The background field is given by Eqs.\,\ref{eq:B}-\ref{eq:E}. In
this simulation , the magnetic ripple field due to the finite toroidal
coils is considered, which is modeled by \cite{RELong_WangYulei2016,laurent_Rax_MagneticRipple1990}
\begin{equation}
\delta\mathbf{B}=\delta B\mathbf{e}_{r}\,,\label{eq:Bripple}
\end{equation}
\begin{equation}
\delta B\left(r,\theta,\varphi\right)=\sum_{m=0,n=1}^{m=\infty,n=\infty}\delta B_{mn}\left(r\right)\mathrm{cos}\left(m\theta\right)\mathrm{cos}\left(nN\varphi\right)\,,\label{eq:MrippleInfty}
\end{equation}
where, $r$, $\theta$, and $\varphi$ are the components of toroidal
coordinate, and $N$ is the number of toroidal coils. Following the
discussion by Laurent \cite{laurent_Rax_MagneticRipple1990}, we consider
only the largest components of the series, $i.e.$, $m=0,\,1$. The
radial profile of magnetic ripple amplitude is approximated by \cite{Ripple_strength_model_Russo_1991}
\begin{equation}
\delta B_{0n}\left(r\right)=\delta B_{1n}\left(r\right)\approx\eta_{n}B_{0}\frac{r^{2}}{a^{2}}\,,\label{eq:deltaB1n}
\end{equation}
where, $\eta_{n}$ are constants denoting the ratio of strength of
$n$th order ripple to $0$th order magnetic strength. Following the
design of ITER \cite{ITER_Design_2002,ITER_Rpple_Portone_2008}, the
tokamak parameters are set as, $B_{0}=5\,\mathrm{T}$, $q=2.5$, $R_{0}=6.2\,\mathrm{m}$,
$a=2\,\mathrm{m}$, $N=18$, and $E_{l}=4\,\mathrm{V/m}$. We consider
the first three harmonics of magnetic ripple. For $n=1$ harmonic,
$\eta_{1}$ is set as $1\%$ as in Ref.\,\cite{ITER_Rpple_Portone_2008}.
For $n=2$ and $n=3$ harmonics, the ratios are given by $\eta_{2}=0.2\%$
and $\eta_{3}=0.06\%$.

The initial poloidal profile of runaway beam is sampled based on a
parabolic distribution, namely,
\begin{equation}
N\left(R,\,z\right)=N_{0}\left[1-\frac{\left(R-R_{0}\right)^{2}+z^{2}}{r_{max}^{2}}\right]\,,\label{eq:N_re}
\end{equation}
where $N_{0}=2.7\times10^{6}\,\mathrm{m}^{-2}$ is the sampling density
at $r=0$, and $r_{max}=1.6\,\mathrm{m}$. The initial toroidal angle,
$\varphi_{0}$, is sampled uniformly from $0$ to $2\pi$; the initial
energy of particles, $\gamma_{0}$ , is sampled based on a normal
distribution with average value $\mu=4.75\,\mathrm{MeV}$ and standard
deviation $\sigma=0.25\,\mathrm{MeV}$; the initial value of pitch-angle,
$p_{\perp0}/p_{0}$, is uniformly sampled from $0$ to $0.3$, and
the initial gyro-phase is sampled uniformly from $0$ to $2\pi$.
The total number of particle samples is on the order of $10^{7}$,
while the total iteration number is approximately $10^{11}$.

\begin{figure}
\includegraphics[scale=0.5]{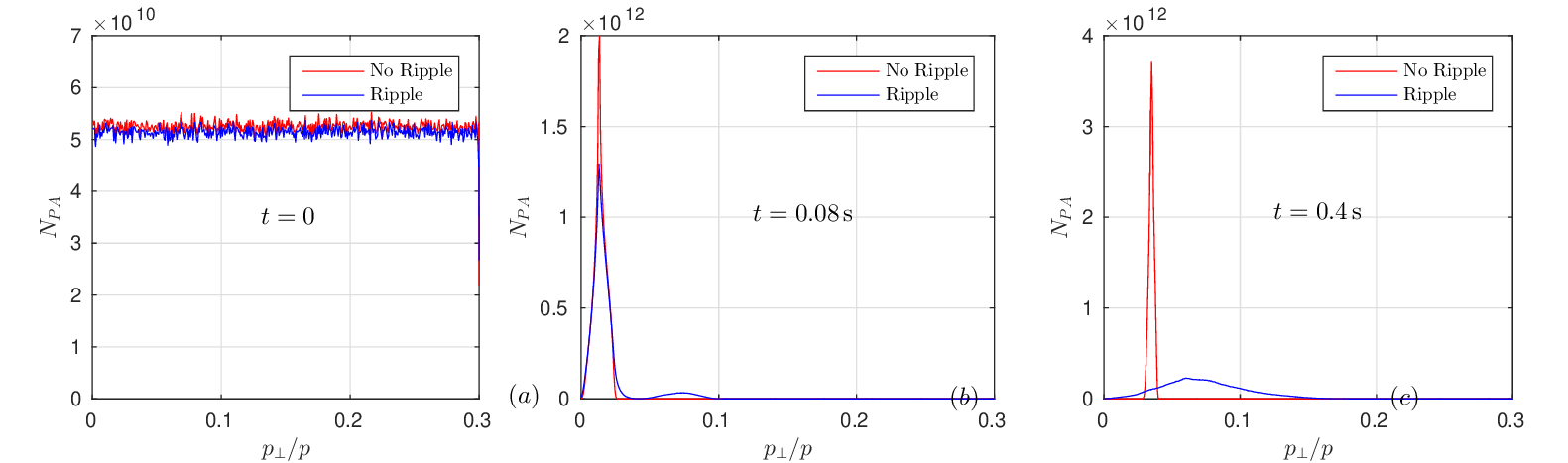}

\caption{The pitch-angle distribution at different moments. The red curves
show the results without magnetic ripple, while the blue curves are
obtained after considering the magnetic ripple field.\label{fig:PA_profile}}
\end{figure}

\begin{figure}
\includegraphics[scale=0.6]{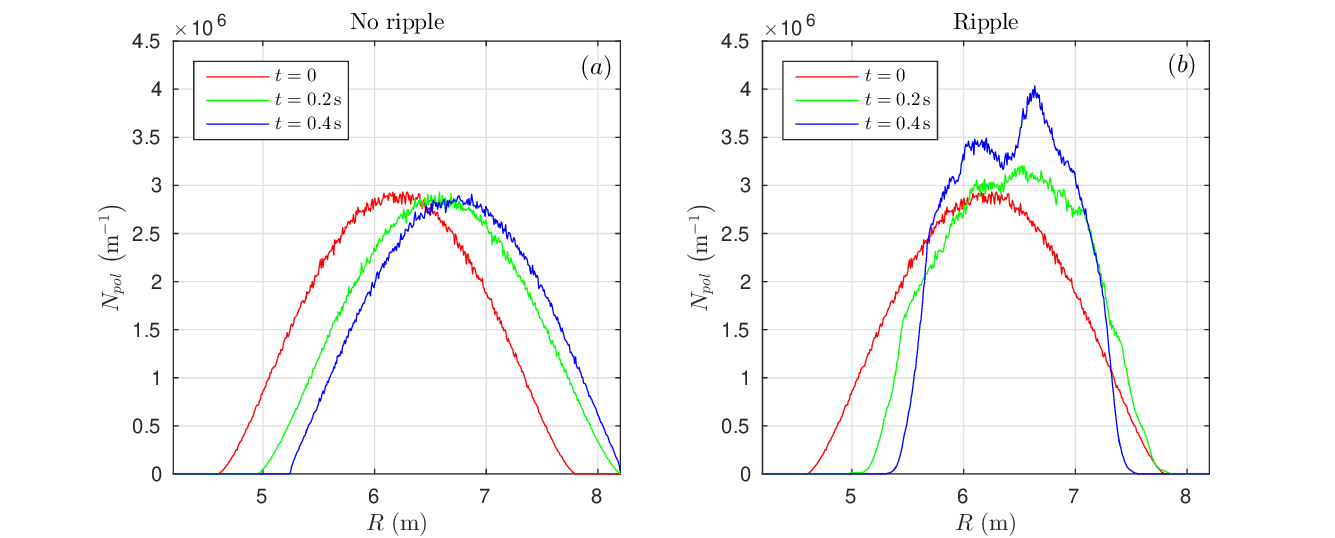}

\caption{The poloidal profile of runaway beam at different moments. (a) shows
the case without ripple field, and (b) depicts the result with magnetic
ripple field. The magnetic ripple can concentrate the runaway profile
near the core of tokamak. \label{fig:N_profile}}
\end{figure}

Figure \ref{fig:PA_profile} depicts the evolution of pitch-angle
scattering with and without the magnetic ripple field. Without magnetic
ripple, the pitch-angle distribution is concentrated in a small interval
after long-term evolution, see the red curves in Fig.\,\ref{fig:PA_profile}a-c.
The pitch-angles of all the particles are approximate $0.04$, which
implies that the initial momentum and position distributions have
little impact on the momentum value at energy limit. However, once
the magnetic ripple field is considered, even though the amplitude
of perturbation is small, the evolution of pitch-angle distribution
shows significant different behavior. To be specific, the pitch-angle
distribution is broadened and the average value of perpendicular momentum
gets larger.

Figure \ref{fig:N_profile} shows the evolution of poloidal profile
of runaway electrons. In the case without ripple field, the runaway
profile drifts outward due to the neoclassical drift \cite{Guan_Qin_Sympletic_RE,Neoclassical_Drift_report},
while the shape of the distribution keeps unchanged. Some electrons
on the right-side hit the wall at $R=R_{0}+a$. After considering
magnetic ripple, the runaway beam is concentrated significantly near
the core of tokamak, and no electron hits the first wall, which implies
a better confinement of the runaways.

\section{Conclusion\label{sec:Conclustion}}

APT provides an efficient platform for large-scale particle simulations
based on geometric algorithms. By tracing sampling particles, APT
can reveal the microscopic and macroscopic dynamical behaviors of
nonlinear and complex systems, such as magnetized plasmas, even in
rather complex geometries. Without extra assumptions, the simulation
model of APT is the first principle model, which obeys the characteristic
line equations of the Vlasov-Maxwell system. The APT code consists
of seven main modules, including the I/O module, the initialization
module, the particle pusher module, the parallelization module, the
field configuration module, the external force-field module, and the
extendible module. The I/O module, supported by Lua and Hdf5 projects,
provides a convenient and flexible interface for both numerical simulation
and data analysis. Under the well-designed integrated and modularized
framework, APT serves as a universal platform for researchers from
different fields, such as plasma physics, accelerator physics, space
science, fusion energy research, computational mathematics, software
engineering, and high-performance computation. 

In this paper, technical details of APT, including the numerical strategy,
the parallelization techniques on different hardware architectures,
and the introductions for using and extending APT, are introduced.
We exhibit the function interfaces of the containers for electromagnetic
fields, external non-electromagnetic force fields, particle pushers,
and initialization approaches. The three-step procedure of extending
APT is provided, and the templates file of extendible functions are
also shown. Following the instructions in Sec.\,\ref{sub:Use-and-extend},
researchers with different backgrounds can use and extend the APT
code conveniently.

A series of new geometric numerical methods and key physical problems,
such as runaway electrons in tokamaks and energetic particles in Van
Allen belt, have been studied using APT. The results reveal the long-term
stability of APT. As an important realization, the APT-SW version
has been successfully distributed on the world's fastest computer,
the Sunway TaihuLight supercomputer, by supporting master-slave architecture
of Sunway many-core processors. Based on large-scale simulations of
runaway electrons under parameters of the ITER tokamak on the Sunway
TaihuLight supercomputer, it is revealed that the magnetic ripple
field can disperse the pitch-angle distribution significantly and
improve the confinement of energetic runaway beam on the same time.
The brand new physical results for solving complex multi-scale dynamical
problems in fusion energy research exhibit the advantages of the APT
code in long-term large-scale applications for complex realistic engineering
problems.

In the next version, the algorithm evaluation module will be supplied
in APT to evaluate the performance of numerical methods under different
complex conditions. Based on this new module, the algorithm recommendation
function can help users to find the optimum algorithm for a specific
problem. The graphical user interface will also be developed to make
APT more convenient to use. APT will also be equipped with more advanced
geometric algorithms and applied to solve more pivotal scientific
problems. 
\ack{
This research is supported by the CAS Key Program of Frontier Sciences
(QYZDB-SSW-SYS004), National Magnetic Confinement Fusion Energy Research
Project (2015GB111003), National Natural Science Foundation of China
(NSFC-11575185, 11575186), JSPS-NRF-NSFC A3 Foresight Program (NSFC-11261140328),
and the GeoAlgorithmic Plasma Simulator (GAPS) Project. 
}

\section*{References}
\bibliography{Refs_RunawayElectrons}

\end{document}